\documentclass[12pt,preprint]{aastex}

\def\LIR{L_{\rm IR}}
\def\Lsun{L_\odot}
\def\Msun{M_\odot}
\def\kms{\,{\rm km\,s^{-1}}}

\def\Gyr{\,{\rm Gyr}}
\newcommand{\myear}{M_\odot\, {\rm yr^{-1}}}

\newcommand{\SFR}{{\rm SFR}}

\begin{document}
\title{Luminous Infrared Galaxies In the Local Universe}
\author{J. L. Wang\altaffilmark{1,2,3},  X. Y. Xia\altaffilmark{3}, 
S. Mao\altaffilmark{4}, C. Cao\altaffilmark{1,2}, Hong Wu\altaffilmark{1},
Z. G. Deng\altaffilmark{5}}
\altaffiltext{1}{National Astronomical Observatories,
                Chinese Academy of Sciences, A20 Datun Road, 100012 Beijing,
                China; Email: wjianl@bao.ac.cn.}
\altaffiltext{2}{Graduate School of the Chinese Academy of Sciences, 100049 Beijing, China.}
\altaffiltext{3}{Dept. of Physics, Tianjin Normal University,
        300074 Tianjin, China.}
\altaffiltext{4}{Jodrell Bank Observatory, University of Manchester,
          Macclesfield, Cheshire SK11 9DL, UK.}
\altaffiltext{5}{College of Physical Science, Graduate School of
        the Chinese Academy of Sciences, 100049 Beijing, China.}
\begin{abstract}
We study the morphology and star formation properties of
159 local luminous infrared galaxy (LIRG) using
multi-color images from Data Release 2 (DR2) of the Sloan Digital Sky Survey (SDSS).
The LIRGs are selected from a cross-correlation analysis between the {\sl IRAS} survey
and SDSS. They are all brighter than 15.9 mag in the $r$-band and below
redshift $\sim 0.1$, and so can be reliably classified
morphologically. We find that the fractions of interacting/merging and spiral galaxies 
are $\sim 48\%$ and $\sim 40\%$ respectively. Our results complement and confirm the
decline (increase) in the fraction of spiral (interacting/merging) galaxies
from $z \sim 1$ to $z \sim 0.1$, as found by Melbourne, Koo \& Le Floc'h (2005).
About 75\% of spiral galaxies in the local LIRGs are barred, indicating that bars 
may play an important role in 
triggering star formation rates $\ga 20\myear$ in the local
universe. Compared with high redshift LIRGs, local LIRGs have
lower specific star formation rates, smaller cold gas fractions
and a narrower range of stellar masses. Local LIRGs appear to be
either merging galaxies forming intermediate mass ellipticals or
spiral galaxies undergoing high star formation activities regulated by bars.
\end{abstract} 

\section{INTRODUCTION}    

Ultraluminous infrared galaxies (ULIRGs) are a class of star forming
galaxies discovered by the InfraRed Astronomical Satellite ({\sl IRAS}). Studies
have convincingly established that they are interacting/merging galaxies
in the local universe (see, e.g., Sanders \& Mirabel 1996 
and Lonsdale et al. 2006 for reviews; see also Cui et al. 2002). These
galaxies are experiencing massive starbursts (with infrared luminosity,
$\LIR>10^{12}\Lsun$) and may eventually 
form sub-$L^*$ early type galaxies (Genzel et al. 2001). There is also
a scaled-up version of ULIRGs at high redshift -- the 
submillimeter emitting galaxies uncovered by deep SCUBA surveys in blank 
fields. These galaxies may, however, be forming more massive galaxies
at earlier times (e.g., Tacconi et al. 2006). 

Recently, luminous infrared galaxies (LIRGs, with $10^{11}\Lsun<\LIR<10^{12}\Lsun$) has
attracted much attention, especially after the launch of the $Spitzer$ $Space$ $Telescope$ (Werner
et al. 2004). The increased sensitivity of $Spitzer$ allows us to probe
this population at high redshift. 
In fact, $Spitzer$ observations of the Chandra Deep Field South (CDFS)
and the Hubble Deep Field North (HDFN) reveal that
the LIRGs contribute more than half of the comoving infrared luminosity
density at $0.5 < z \la 1$ (P\'erez-Gonz\'alez et al. 2005; Le Floc'h et
al. 2005). An important question naturally arises: what is the morphology of these
LIRGs? Are they mostly merging galaxies, just like ULIRGs as pointed out by Arribas et al. (2004)?
The answer may offer clues how the high star formation rate in these galaxies is triggered.

The study of LIRGs is also important for understanding the star
formation history in the universe.
It is well known that the comoving star formation density steeply declines 
from $z\sim 1$ to 0 by a factor of $\sim 10$ (e.g., Lilly et al. 1996; Madau et al. 1998). 
Recent $Spitzer$ observations of the CDFS and HDFN
confirm the rapid decline of the comoving infrared luminosity since
redshift $\sim 1$ (P\'erez-Gonz\'alez et al. 2005; Le Floc'h et al. 2005). 
The rapid decline in the star formation rate does not, however, mean that the stellar
mass assembled is negligible since redshift $\sim 1$.
From the spectral energy distributions of galaxies, Dickinson et al. (2003) 
concluded that 30\% to 50\% of the stellar mass in present-day galaxies
has been formed since $z \sim 1$. Most of this stellar mass buildup
cannot occur in massive early-type galaxies as most of their stellar mass is
already in place by redshift $\sim 2$  (e.g., Reddy et al. 2006). 
So the substantial stellar mass assembly since redshift $\sim 1$
likely occurs in intermediate-mass galaxies (so called ``downsizing'',
Cowie et al. 1996; Hammer et al. 2005; Bell et al. 2005; Juneau et al. 2005), with mass between
${\rm few}\times 10^{10} \Msun$ to ${\rm few} \times 10^{11}\Msun$.
As LIRGs contribute the majority of the star formation rate at
$z\sim 1$, it is important to understand how this population 
evolves as a function of redshift in order to understand the decline
in the star formation rate from redshift 1 to the local universe.

Zheng et al. (2004) studied the morphologies of 36 distant LIRGs ($0.4<z<1.2$) with HST images.
They find that only 17\% LIRGs are obvious mergers and the
fraction of interacting/merging systems is at most 58\%. Furthermore,
$\sim 36\%$ distant LIRGs are classified as normal disk galaxies and
$\sim 25\%$ are compact sources. The average stellar mass of LIRGs is $\sim 10^{11}\Msun$. 
This study shows that a large fraction of distant LIRGs are star forming disk 
galaxies through various morphological phases at
intermediate redshift, and a substantial fraction of the stellar mass in disk galaxies 
may be assembled through an LIRG phase since redshift 1. Although the
sample size of Zheng et al. (2004) is small, subsequent works based on
larger samples arrived at similar classification conclusions. For example, Bell et al. (2005) 
investigated several hundred infrared luminous galaxies with 
$5\times10^{10}\Lsun<\LIR<3\times10^{11}\Lsun$ at z $\sim 0.7$ in CDFS and find that more than
half are massive spirals and less than a third are strongly interacting/merging galaxies. 

Melbourne et al. (2005) performed the most comprehensive study on the
morphological evolution of LIRGs since $ z\sim  1$ using 119 LIRGs in 
the Great Observatories Origins Deep Survey-North field (GOODS-N). Their
sample covers the redshift range from $z \sim 0.1$ to 1. They derived the
optical morphologies and photometries from HST images, redshifts from
Keck observations, and the infrared luminosities from $Spitzer$ observations. 
They also find evidence for the morphological evolution for LIRGs in the 
last $\sim $ 8 Gyr. Above redshift 0.5, about half of LIRGs are spirals and the 
ratio of peculiar/irregular to spiral is about 0.7, and
all morphological classes of LIRGs span a similar range of infrared and 
optical luminosities. In contrast, at lower redshift, spirals account
for just one third of LIRGs, and they also appear to be slightly
fainter than peculiar/irregular galaxies at similar redshift.

In addition to the morphological evolution, LIRGs in the local universe
and those at high redshift may also differ in other aspects.
For example, Reddy et al. (2006) find, from deep $Spitzer$ MIPS 24
$\mu$m observations that LIRGs at $z \sim 2$ have a wide range of stellar mass, spanning
from $2\times 10^{9} \Msun$ to $5 \times 10^{11}\Msun$. Furthermore,
based on $Spitzer$ $24\mu{\rm m}$ observations,
P\'erez-Gonz\'alez et al. (2005) investigated
the evolution of the specific star formation rate (SFR per unit stellar mass) 
as a function of the total stellar mass since $z\sim 3$ and find that LIRGs
at higher redshift have higher specific SFRs than the local counterparts.
These differences hint that local LIRGs may not be the exact analogues
of LIRGs at high redshift, and thus it is important to establish
the morphology and star formation properties of  local LIRGs.

So far the local sample used to establish the evolution 
trend is still small. Ishida (2004) performed a morphological
classification for a low redshift sample with 56 LIRGs drawn from the
{\sl IRAS} Bright Galaxy Sample (Soifer et al. 1987). Clearly a larger 
sample is desirable to firmly establish the properties of the local
LIRGs, including their morphologies and star formation properties.
We assemble such a sample by cross-correlating the {\sl IRAS} source catalogue
and the Data Release 2 (DR2) of the Sloan Digital Sky Survey (SDSS, see below).
Our LIRG sample includes 159 objects with redshift $z \la 0.1$.
Due to the low-redshift of our objects, the linear physical resolution  
of the SDSS images is comparable to that achieved for the high redshift 
objects observed with ACS on board the HST  (Melbourne et al. 2005). 
Therefore comparisons in the morphology of our local sample and those 
at high redshift can provide a clear picture for the morphological evolution 
of LIRGs since $z \sim 1$. 
The outline of the paper is as follows. In \S2 we describe our local LIRG
sample and in \S3 we discuss how we perform the morphological
classification and derive the star formation properties in LIRGs.
We present our main results in \S4 and finish with conclusions in \S5.
Throughout this paper we adopt a cosmology with
a matter density parameter $\Omega_{\rm m}=0.3$, a cosmological constant
$\Omega_{\rm \Lambda}=0.7$ and
a Hubble constant of ${\rm H}_{\rm 0}=70\,{\rm km \, s^{-1} Mpc^{-1}}$.

\section{THE SAMPLE \label{sec:sample}}

Our sample LIRGs are drawn from the LIRG catalog of Cao et al. (2006),
who carried out a cross-correlation study of the {\sl{IRAS}} point
source catalogue (PSC) and faint source catalogue (FSC) (Moshir et al. 1992) 
with DR2 of the SDSS (Abazajian et al. 2004).  The total
number of LIRGs identified with high reliability from the FSC is 908. 
The {\sl IRAS} survey covers all sky while the DR2 covers only 2627 
square degree of sky at a (Petrosian) magnitude limit of 17.77 mag in 
the $r$ band. Note that the catalog of Cao et al. (2006) may miss some 
local LIRGs that are too faint (for example, due to heavy dust extinctions) 
to be included in the SDSS (for detail see Cao et al. 2006). 

As pointed out by Fukugita et al. (2004), a reliable visual morphological
classification can be performed only for SDSS galaxies brighter than $r=15.9$ mag
after correcting for the Galactic extinction (Schlegel et al. 1998). 
In addition, SDSS spectroscopic selection of galaxies become incomplete at 
$r<14.5$ mag (Kauffmann et al. 2003).
We therefore restrict our sample LIRGs to be $14.5<r<15.9$, 
which reduces the number of LIRGs to 210. 
Furthermore, we selected only the LIRGs from the FSC LIRG
sample of Cao et al. (2006) with {\sl IRAS} 60$\mu$m flux greater than
0.3 Jy (rather than $\sim$ 0.2 Jy for the whole sample) to maximize
the sample completeness. This step further
reduces the number of LIRGs to 159. While our selection criteria
remove half of LIRGs with $z>0.1$, we show in
\S\ref{sec:result} below that these do not lead to significant
biases in morphological classifications and stellar mass distributions.

The {\sl{r}}-band magnitude, redshift and infrared luminosity distributions of 
the sample are shown in Fig. 1. It is clear from Fig. 1 that our sample LIRGs
have $z\la 0.1$, and most are in the 
redshift range of 0.04 $< z <$ 0.08. The LIRG sample in this redshift
range is just what is needed to extend the analysis for the evolution of
LIRGs based on CDFS and GOODS-N from high redshift to the local
universe. Fig. 1 (bottom left panel) also shows that
most of our LIRGs have infrared luminosity smaller than $\sim 6\times10^{11} L_{\odot}$. 

\section{MORPHOLOGICAL CLASSIFICATION AND ESTIMATION OF STAR FORMATION PARAMETERS}

In this section, we first describe our classification schemes
and procedures in detail in \S\ref{sec:morph}, and then discuss how we derive the stellar mass, star
formation rate, and cold molecular gas mass in \S\ref{sec:SFR}. For the majority
of our galaxies ($\sim 73\%$), the star formation properties can be derived in a
straightforward manner, however for some galaxies (e.g., those in a pair or group of
galaxies, see Fig. 2 for examples), extra care must be taken, and this is described in more detail
in \S\ref{sec:extra}.

\subsection{Morphological Classification \label{sec:morph}}

  In order to compare the morphology of local LIRGs and those at high redshift, 
we adopted the same classification scheme as Melbourne
et al. (2005), who divided galaxies into peculiars (interacting and
merging), spirals (including barred and non-barred, face-on and edge-on
spirals), compact galaxies and elliptical galaxies.
The peculiar galaxy type is characterized by irregular, asymmetric shape 
and show clear interacting/merging sign or merger relics, such as tidal
features. The compact type is signified by their small sizes (for a
quantitative criterion, see below), while spirals and ellipticals have
their usual definitions, for example the spirals
have (largely) symmetric shape, obvious disk and bulge components
and exhibit no disturbed spiral arms.

  We performed the visual classification for all LIRGs primarily in the
{\sl{r}}-band, but also used composite color images as a
cross-check, a procedure adopted by Zheng et al. (2004) and Melbourne et al. (2005) as well.
The color images were produced by combining ($g$, $r$, $i$) filter data following
Lupton et al. (2004)\footnote{All the images are available at
http://www.jb.man.ac.uk/\~\,smao/LIRG.html}. The classification was done
independently by four of us (JLW, XYX, ZGD, HW). For the majority of galaxies, the
classification was the same. However, for $\sim 20\%$ of LIRGs, the
classifications differ. For these, a consensus classification was adopted after discussions.

In light of the important
role galactic bars may play in star formation processes, we further
divided the spirals into barred and non-barred types. Reliable visual identification of
bars can, however, only be performed for low-inclination angle ($i$) 
spirals with $i < 60^{\circ}$ ($i=90^\circ$ means edge-on).
Furthermore only strong/obvious bars can be identified with high confidence. 
So to double check the reliability of visual bar identifications,
we also used a quantitative method to identify bars by examining the changes
of isophote shapes in the $r$-band and the color images (e.g., Jogee et
al. 2004; Zheng et al. 2005). This method relies on the fact that
at the end of strong bars, there are often rapid changes in the position angle
and ellipticity. This approach identified nearly the same
barred spirals as the visual search, so we believe our classification of
barred galaxies is reliable.

In practice we first selected isolated compact galaxies using the criterion that
their half-light radii must be smaller than 3 kpc. The half-light radius
is taken to be the $r$-band Petrosian radius $R_{50}$, within which half of
the $r$-band Petrosian luminosity is enclosed. There are 19 isolated
compact galaxies in our LIRG sample. For the remaining 139 LIRGs, we
carried out the morphological classifications as discussed above. Note also that
there are no early type galaxies in our sample, so our objects fall into
only three categories (peculiar, spiral, and compact galaxies).

\subsection{Stellar Mass, Star Formation Rate, and Cold Molecular Gas Mass \label{sec:SFR}}

LIRGs are undergoing high star formation activities, and so the stellar
population in these galaxies is a mix of young and old stars.
The stellar mass for the old (evolved) stellar population is estimated based
on the SDSS photometric data following Bell et al. (2003): 
\begin{equation}
\log (M_*/M_{\odot}) = -0.4(M_{\rm r, AB} - 4.67)+[a_{\rm r} + b_{\rm r}
\times (g-r)_{\rm AB} + 0.15],
\end{equation}
where $M_{\rm r,AB}$ is the $r$-band absolute magnitude, $(g-r)_{\rm AB}$ is the 
rest-frame color in the AB magnitude system. The term 4.67 is the absolute magnitude of 
the Sun in the SDSS $r_{\rm AB}$ band. The photometric $k$-correction is calculated using
the method of Blanton et al. (2003, {\tt kcorrect} v4\_1\_4). The coefficients $a_{\rm r}$ 
and $b_{\rm r}$ are taken from Table 7 of Bell et al. (2003). 
Throughout this paper a Salpeter (1955) stellar initial mass function (IMF)
is used with $dN/dM \propto M^{-2.35}$ and $0.1M_{\odot}<M<100M_{\odot}$.
This gives a stellar mass 0.15 dex higher (the last term in eq. 1) than 
that in Bell et al. (2003) where a ``diet''  Salpeter stellar IMF is used. 
The stellar mass estimated using eq. (1) is uncertain by about
0.3\, dex due to the effects of dust, galaxy age and star formation history.
The bottom right panel of Fig. 1 shows the stellar mass distribution of
159 local LIRGs. A more detailed discussion will be presented in
\S\ref{sec:SFR2}, together with their velocity dispersion information.

The star formation rate (SFR) of LIRGs is derived following Kennicutt (1998a)
\begin{equation}
\SFR =4.5 \,\myear {L_{\rm IR} \over 10^{44}\, {\rm erg\,s^{-1}}},
\end{equation}
where $L_{\rm IR}$ is the infrared luminosity between 8-1000\,$\mu$m;
the systematic uncertainty in the SFR normalisation is about a factor
of $\sim $ 2-3 (Kennicutt 1998b). The bottom left panel of Fig. 1 shows the 
distribution of SFR for our LIRG sample.

The cold molecular hydrogen reservoir in LIRGs will limit how
many new stars can be formed in the current starburst episode, and so it
is important to estimate the molecular hydrogen mass. To do this, 
we first derive the CO luminosity using the correlation between 
the CO luminosity and the infrared luminosity obtained by Gao \& Solomon (2004a),
\begin{equation}
\log L_{\rm IR}=1.27 \log L_{\rm CO}' - 0.85,
\end{equation}
and then obtain the molecular hydrogen mass using the relation
\begin{equation}
M(H_2)=\alpha  L_{\rm CO}'
\end{equation}
(Gao \& Solomon 2004b; Solomon \& Bout 2005), where we take $\alpha$ as
4.6 $M_{\odot}~(\rm K~km~s^{-1}~pc^2)^{-1}$, the approximate value for the Milky
Way. Note that the adopted conversion factor $\alpha$ 
may over-estimate the molecular gas mass by
a factor of few for interacting/merging LIRGs (Solomon \& Bout 2005).

\subsection{Parameter Estimation for Merging and Deblended Galaxies \label{sec:extra}}

  For the majority ($\sim 73\%$) of our galaxies, the procedure to
derive various quantities as discussed
in the last section is straightforward. However, for 
some ($\sim 27\%$) of our galaxies, the determination is not simple for
two reasons. First, given the poor IRAS resolution (the
average major axis of the $2\sigma$ error ellipse in position is approximately
40$\arcsec$ at 60 $\mu$m), there may be a pair or a group of galaxies
within an IRAS $2\sigma$ error ellipse. Fig. 2 shows three examples. In such cases,
it is difficult to determine which galaxy (or galaxies) is the true contributor
of the infrared emission. Second, the SDSS photometric pipeline tends to
deblend very extended galaxies or remnants of mergers into several small
parts. For multi-nucleus systems, the SDSS pipeline may also
identify them as several galaxies. When we performed the 
cross-correlation of an IRAS object with the SDSS catalogue,
only one optical object in the SDSS is identified as the counterpart of
an IRAS source. For artificially deblended objects, this
will result in an under-estimate of the stellar
mass for artificially deblended objects. 

  To solve the problem of deblending, we carefully examined the
SDSS image of each object. We flag all objects within 3 Petrosian half-light radii ($R_{50}$) 
that are brighter than 10\% of the identified galaxy in either the $g$- and $r$-band.
After excluding probable foreground and background objects, the stellar mass of the identified galaxy and 
all the deblended parts are summed up to give the total stellar mass. 
Galaxies outside $3R_{50}$, but with obvious
merger signs (such as tidal tails and bridges) are also included in the
total stellar mass.

    As we mentioned above, due to the low spatial resolution of IRAS, when the identified source is
in a pair or group of galaxies, the infrared luminosity may be from the whole
system, rather than just from a single object. For these sources, we visually
examined the region covered by the IRAS $2\sigma$ error ellipses in the
$r$-band images. We include all galaxies with obvious interacting/merging
signatures in the infrared emission and the stellar mass calculation. 
For those galaxies without obvious signs of interacting or 
mergers, we also checked whether their redshifts
are available in the SDSS or the NASA/IPAC Extragalactic Database (NED) 
\footnote{The NASA/IPAC Extragalactic Database (NED) is operated by the
Jet Propulsion Laboratory, California Institute of Technology, under
contract with the National Aeronautics and Space Administration.}.
If they are and their redshifts are the the same as our target, then we also include them
as a contributor of the infrared emission. For F12229-0233 in Fig. 2,
there are three galaxies within the IRAS $2\sigma$ error ellipse, so we
included all of them in the stellar mass and infrared luminosity calculations.
Fog NGC 5331 and UGC 11673, the error ellipse intercepts one 
member of an interacting pair. For these two cases, we 
included both in the total stellar mass budget. For NGC 5331,
this is consistent with the high-resolution images obtained
by Surace et al. (2004) who presented a re-analysis of IRAS
images for 106 interacting galaxies. The infrared contours for NGC 5331
(at 10\% peak flux level, see their Fig. 1) from
12\,$\mu$m to 100 $\mu$m clearly encloses both galaxies, 
suggesting that we should include both galaxies in the infrared luminosity
and stellar mass calculations.

  In total, there are 43 optical counterparts of IRAS sources
where the stellar mass is calculated from the deblended or multiple components 
of the system. Among them, 23 objects have companions or deblended objects
occurred within  $3 R_{50}$, and 23 have companions outside. There are
three (overlapping) cases where there are
deblended components within $3 R_{50}$, and interacting pairs outside 
$3 R_{50}$.

  Our procedures are not perfect. The first problem
arises due to un-identified contributors to the infrared emission.  Companions
that contribute to the infrared emission, but without obvious merging
signatures, will be missed in our stellar mass budget. The second
problem is that if the infrared emission comes from only one galaxy in
an interacting or merging system, our procedure 
will obviously over-estimate the stellar mass corresponding to the infrared galaxy,  
as we include the stellar mass from all members. 
Such ambiguous cases are relatively few in number, and so our
results should not be significantly affected, although we caution that
interacting/merging galaxies will suffer most from the problems discussed above.

\section{RESULTS \label{sec:result}}

\subsection{Morphology of Local LIRGs and Comparisons with high redshift}

Out of the 159 sample LIRGs, 19 LIRGs are classified as isolated compact
category according to the criterion that their half-light radii should
be smaller than 3 kpc. 
77 (48\%) can be unambiguously classified as interacting/merging 
galaxies or merger remnants with obvious merger relics. There are
an additional 9 objects which show signs of possible interacting or relics of merger. 
If we add them into the interacting/merging class, the total fraction
can be as high as $\sim 54\%$.
This fraction is still significantly smaller than that in ULIRGs where almost all are 
merging/interacting systems. For the remaining LIRGs, 63 (40\%) are classified as
normal spirals, of which 11 have high inclination angles ($i > 60^{\circ}$) 
and hence difficult to identify bars in these galaxies.
For the remaining 52 spirals with low inclination angles, 39 (75\%) show
clear strong bars. This large fraction of barred spirals indicates that
bars may play an important role in triggering the intense star formation activities in the LIRGs.

Fig. 2 shows three examples for each category. The first and second columns 
show the normal spirals and barred spirals respectively. 
Clearly some barred spirals have close companions (at least in projection). 
It is possible that such minor tidal interactions produce both the bar and 
the symmetric spiral arms, and provide the perturbations to channel 
the gas to the central regions. The third and fourth columns of Fig. 2 show 
the interacting and merging LIRGs, which exhibit all possible interacting/merging 
stages, such as interacting pair galaxies, multi-merging groups of galaxies, 
mergers with two close nuclei, and galaxies with a single nucleus but with
merger relics. The compact class of LIRGs is shown in the last column of
Fig. 2. They are on average at smaller redshift than the whole sample.
These objects all have compact sizes, but their morphologies 
appear to be diverse, ranging from spirals with clear symmetric arms to merging galaxies with 
tidal features, implying that they may not have a homogeneous physical origin.
From the spectroscopic information (Brinchmann et al. 2004), 
the active galactic nuclei (AGN) fraction in the compact category is not much 
higher than that in other classes of LIRGs, and so AGNs may not be the main
reason for their compact morphology.

Our morphological classifications for local LIRGs are plotted in Fig. 3 together
with the corresponding results from those of LIRGs at higher redshift ($0.1<z<1$) taken 
from Table 1 of Melbourne et al. (2005). Our results are consistent 
with the extrapolation of theirs to the local universe. These results 
clearly indicate that the fraction of interacting/merging galaxies increases 
from about $\sim 30\%$ at $z \sim 1$ to about $50\%$ at  $z \sim 0$,
accompanied a decrease in the fraction of spirals from roughly 50\% to
40\%. Note that most of the changes occur between redshift 1 and 0.3 and
there are virtually no changes from redshift 0.3 to the present day.

As discussed in \S\ref{sec:sample}, to obtain reliable visual
classifications, we only selected galaxies brighter than 15.9 in the $r$-band.
To see how this magnitude cutoff affects the morphological mixes,
we also performed rough classifications for 
381 LIRGs in the FSC sample of Cao et al. (2006) 
with $15.9<r<17.77$ mag and below redshift 0.1.
We find that the relative fractions of different categories do not
differ significantly from those of the bright LIRG sample ($r<15.9$) we
focused on. Thus our magnitude cutoff 
has not introduced significant biases into the morphological types of local LIRGs. 
Note also that the LIRG samples used by Reddy et al. (2006) and Le Floc'h et al. (2005)
are also optical magnitude and infrared flux limited samples.

\subsection{Stellar Mass, Specific SFR and Cold Molecular
Mass of Local LIRGs \label{sec:SFR2}}

  The bottom right panel of Fig. 1 shows the stellar mass distribution of 159 local LIRGs.
We can see from Fig. 1 that the stellar mass of the old stellar
population for most local LIRGs is in the range of
3$\times$10$^{10} M_{\odot}$ to 5$\times$10$^{11} L_{\odot}$,
indicating most local LIRGs are intermediate mass galaxies, 
within a factor of few of the stellar mass in the Milky Way
(where $M_* \sim 7 \times 10^{10} M_\odot$, see Mo \& Mao 2004, \S2.3). 

The left column of Fig. 4 further examines the distribution of infrared luminosity 
for each category of LIRGs. As can be seen, the peculiar galaxies appear 
to have a relatively extended tail
toward larger infrared luminosity compared with other types, while the
compact type appears to be skewed toward smaller values.
Table 1 lists in more detail the number of objects for each morphological 
class in different infrared luminosity ranges. From Table 1, we can see that 
the fraction of interacting/merging LIRGs increases from 40\% to 100\% when the 
infrared luminosity increases from $10^{11}\Lsun$ to larger than $4\times10^{11}\Lsun$. 
In contrast, there are no spirals with infrared luminosity larger than $4\times10^{11}\Lsun$,
implying that the local spirals LIRGs have lower infrared luminosities than 
those of interacting/merging LIRGs. The latter dominates in number at higher infrared
luminosity, just as in ULIRGs. This result is consistent with that
obtained from a deep wide-field survey for LIRGs by Ishida (2004).

The middle column in Fig. 4 illustrates the stellar mass distribution for each category of LIRGs.
Although the shapes of distributions differ in detail for different morphological 
types, the median stellar mass is almost the same for normal disk galaxies and 
interacting/merging system; only the compact galaxies has a median stellar 
mass that is lower by about 0.2 dex. 
We also checked the stellar mass distributions for the whole LIRG sample of 
Cao et al. (2006) with $ 14.5 < r < 17.77 $, rather than only for
the 159 galaxies with $r<15.9$ as shown in Fig. 4. The stellar mass
range of the fainter sample is similar to the brighter one, suggesting
that our magnitude limit (imposed in order to obtain reliable visual
morphological classifications) had no significant impact on the stellar 
mass distribution. Therefore, the stellar mass range of local
LIRGs appears to be narrower than that of high redshift LIRGs 
(from $\sim 2\times 10^{9}\Msun$ to $6 \times 10^{11}\Msun$), as found by
several recent studies using {\it Spitzer} 24$\mu$m observations (Reddy
et al. 2006; Le Floc'h et al. 2005; P\'erez-Gonz\'alez et al. 2005).

The stellar mass inferred using eq. (1) depends on various assumptions
such as stellar populations. It is therefore interesting to cross-check whether their dynamical
properties are also consistent with those of an intermediate mass population.
The MPA group provides estimates of the velocity dispersion of these
galaxies (D. Schlegel, et al. in preparation as quoted in Heckman et
al. 2004). There are 152 our LIRGs with available 
velocity dispersions \footnote{http://www.mpa-garching.mpg.de/SDSS/} and 
have a median per-pixel signal-to-noise ratio greater than 10.
As the instrumental resolution of 
the SDSS is approximate 70 km s$^{-1}$, we exclude objects with velocity 
dispersion smaller than 70 km s$^{-1}$.  Visual checks of the
images also reveal that three objects have spectroscopic locations
displaced from the central regions of galaxies, and so they are
excluded from the velocity dispersion analysis, leaving us with a total of
147 LIRGs with velocity dispersions. The right column of Fig. 4 shows
the histograms of velocity dispersion for different types of galaxies.
It can be seen that all types of galaxies have similar ranges
of velocity dispersions; the median value for each type is  
$\sim 150\kms$, similar to that ($\log \sigma_*=2.220$) for 
an $L^*$ galaxy (see Table 1 Bernardi et al. 2005). Note that the velocity dispersions for several 
objects in the peculiar class exceed $300\kms$, which indicates that these
galaxies may not have reached dynamical equilibrium.

There may be other differences between the local and high redshift
LIRGs as well. The left panel of Fig. 5 shows the specific SFR (SFR per units stellar mass)
vs. the stellar mass for sample LIRGs. Clearly,
there is a trend that more massive LIRGs have smaller specific
SFRs, which is also seen in studies of high redshifts LIRGs based on $Spitzer$ observations.
This trend is partly due to the inverse proportionality of the specific SFR on
the stellar mass. Quantitatively, Fig. 5 shows that the SFRs
for most of sample LIRGs are between 10 to 100 $\Msun\,{\rm yr}^{-1}$,
with a median of about 30 $\Msun\,{\rm yr}^{-1}$. Most of the specific SFRs lie between 0.06 $\Gyr^{-1}$ to 
$1\Gyr^{-1}$,  consistent with those derived by 
P\'erez-Gonz\'alez et al. (2005) for LIRGs below redshift 0.4 (see their
Fig. 11). Our sample extends the {\it Spitzer} samples to the local
universe, as they have very few galaxies below redshift $0.2$. In
contrast, the specific SFRs of high redshift LIRGs
span two order of magnitude from $1\Gyr^{-1}$ to $ \sim 100 \Gyr^{-1}$
(P\'erez-Gonz\'alez et al. 2005), substantially higher than those for the local LIRGs.

As shown in \S\ref{sec:SFR}, the stellar mass and cold molecular gas mass 
can be estimated using eqs. (1-4), which can then be combined to 
derive the cold gas fraction, $f_{\rm cold}=M(H_2)/(M(H_2)+M_*)$. This
is shown in the right panel of Fig. 5. The average cold gas fraction 
is $\sim 10\%$, which is substantially smaller than that at high redshift, 
namely about $\sim 50\%$ (Reddy et al. 2006). 
The low gas fraction ($\sim 10\%$) in the local LIRGs also implies that
the total stellar mass built in the LIRGs episode will be limited, i.e.,
the stellar mass buildup has almost ceased in the local universe. 
Note that if the conversion factor $\alpha$ adopted in eq. (4)
over-estimates the molecular gas mass 
in the local universe for interacting/merging LIRGs (Solomon \& Bout
2005), the difference in the cold gas mass fraction
between low- and high-redshift will be even larger.

\section{CONCLUSIONS}

In this paper, we have studied a sample of 159 LIRGs obtained by cross-correlating
IRAS and the DR2 of the SDSS. We performed a careful morphological
classification of these galaxies and studied their physical
properties. We compared their properties with higher redshift LIRGs, and
found systematic differences. Our main conclusions are as follows:
\begin{enumerate}
\item In the local LIRGs, the fractions of interacting/merging and spiral galaxies 
are $\sim 48\%$ and $\sim 40\%$ respectively. This confirms the decline
of the fraction of spiral galaxies in LIRGs
from redshift 1 to the local universe, as first found by Melbourne, Koo \& Le Floc'h (2005).
\item The majority ($\sim 75\%$) of our disk galaxies are strongly
barred, indicating that bars may play a key role in 
star forming galaxies (with SFR $\ga 20\myear$) in the local
universe.
\item Most LIRGs are intermediate mass galaxies, as seen from their
inferred stellar masses and observed velocity dispersions.
\item Compared with high redshift LIRGs, local LIRGs have
lower specific star formation rates, smaller cold gas fractions
and a narrower range of stellar masses.
\end{enumerate}

We conclude that most local LIRGs are either 
major mergers between gas-rich spirals enroute to the formation of 
intermediate mass ellipticals, or massive disk galaxies undergoing an episode 
of star formation regulated by bars. Most of the gas in the local
universe has already been turned into stars, and hence 
gas-rich galaxies are rare. Either major mergers and/or bars are required 
to bring the gas into the centers of galaxies and trigger the high star 
formation rates seen in LIRGs (see Jogee, Scoville \& Kenney 2005 for an 
excellent discussion). On the other hand, higher redshift
LIRGs have much higher gas fractions (Reddy et al. 2006). The higher gas
reservoir makes it more likely that the LIRG phase can occur in more
numerous galaxies, and in a variety of galaxies 
ranging from dwarf galaxies, galactic bulges to cores of 
ellipticals, resulting in a larger range of stellar masses than local
LIRGs.

While the local LIRGs are rare, and can just account for about 5\% of the 
total infrared energy from galaxies (Le Floc'h et al. 2005), nevertheless, 
their proximity renders detailed studies of these objects easier than their 
much more abundant high redshift counterparts. The understanding of these 
objects may provide important insights into the physical mechanisms that 
drive the star formation process in the local universe, and how disks and 
spheroidals are assembled both locally and in the past.

\acknowledgements

We thank X. Z. Zheng, C. N. Hao and Z. L. Zou for advice and helpful discussions.
Thanks are also due to the anonymous referee for constructive comments.
This project is supported by NSF of China No.10333060, No.10473013 and No.10640430201.
XYX acknowledges financial support by the visitor's grant at
Jodrell Bank. SM thanks the Chinese Academy of Sciences and Tianjin
Normal University for travel support.
Funding for the creation and distribution of the SDSS 
Archive has been provided by the Alfred P. Sloan Foundation, the 
Participating Institutions, the National Aeronautics and Space Administration, 
the National Science Foundation, the U.S. Department of Energy, the Japanese 
Monbukagakusho, and the Max Planck Society. The SDSS Web site is http://www.sdss.org/. 
The SDSS is managed by the Astrophysical Research Consortium (ARC) for the 
Participating Institutions. The Participating Institutions are The University 
of Chicago, Fermilab, the Institute for Advanced Study, the Japan Participation 
Group, The Johns Hopkins University, the Korean Scientist Group, Los Alamos 
National Laboratory, the Max-Planck-Institute for Astronomy (MPIA), the 
Max-Planck-Institute for Astrophysics (MPA), New Mexico State University, 
University of Pittsburgh, Princeton University, the United States Naval Observatory, 
and the University of Washington.

\clearpage

\begin{deluxetable}{rcc|cc|cc|cc}
\tabletypesize{\scriptsize}
\tablewidth{0pt}
\tablecaption{Morphology classification of LIRGS as a function of infrared luminosity}
\tablehead{
\colhead{} &
\multicolumn{2}{l|}{11 $<$ log $\LIR/L_{\odot}$ } &
\multicolumn{2}{l|}{11$<$ log $\LIR/L_{\odot}$ $<$ 11.3} &
\multicolumn{2}{l|}{11.3 $<$ log $\LIR/L_{\odot}$ $<$ 11.6} &
\multicolumn{2}{l}{11.6 $<$ log $\LIR/L_{\odot}$ }
}
\startdata
                &   N       & \%        &   N      & \%     & N      &  
\%     &N      &  \%    \\
Total           & 159       &100        & 111      &100     &39      
&100      &9      &100     \\
Non-barred spiral& 13        &8          & 10       &9       &3       &8        
&0      &0        \\
Barred spiral   & 39        &25         & 30       &27      &9       
&23       &0      &0        \\
Highly-inclined & 11        & 7         & 10       & 9      &1       &3        
&0      &0         \\Peculiar        & 77        &48         & 45       
&41      &23      &59       &9      &100        \\Compact         & 19        
&12         & 16       &14      &3       &8        &0      &0        
\\\enddata
\end{deluxetable}

\clearpage
\begin{figure}
\figurenum{1}
\epsscale{}
\plotone{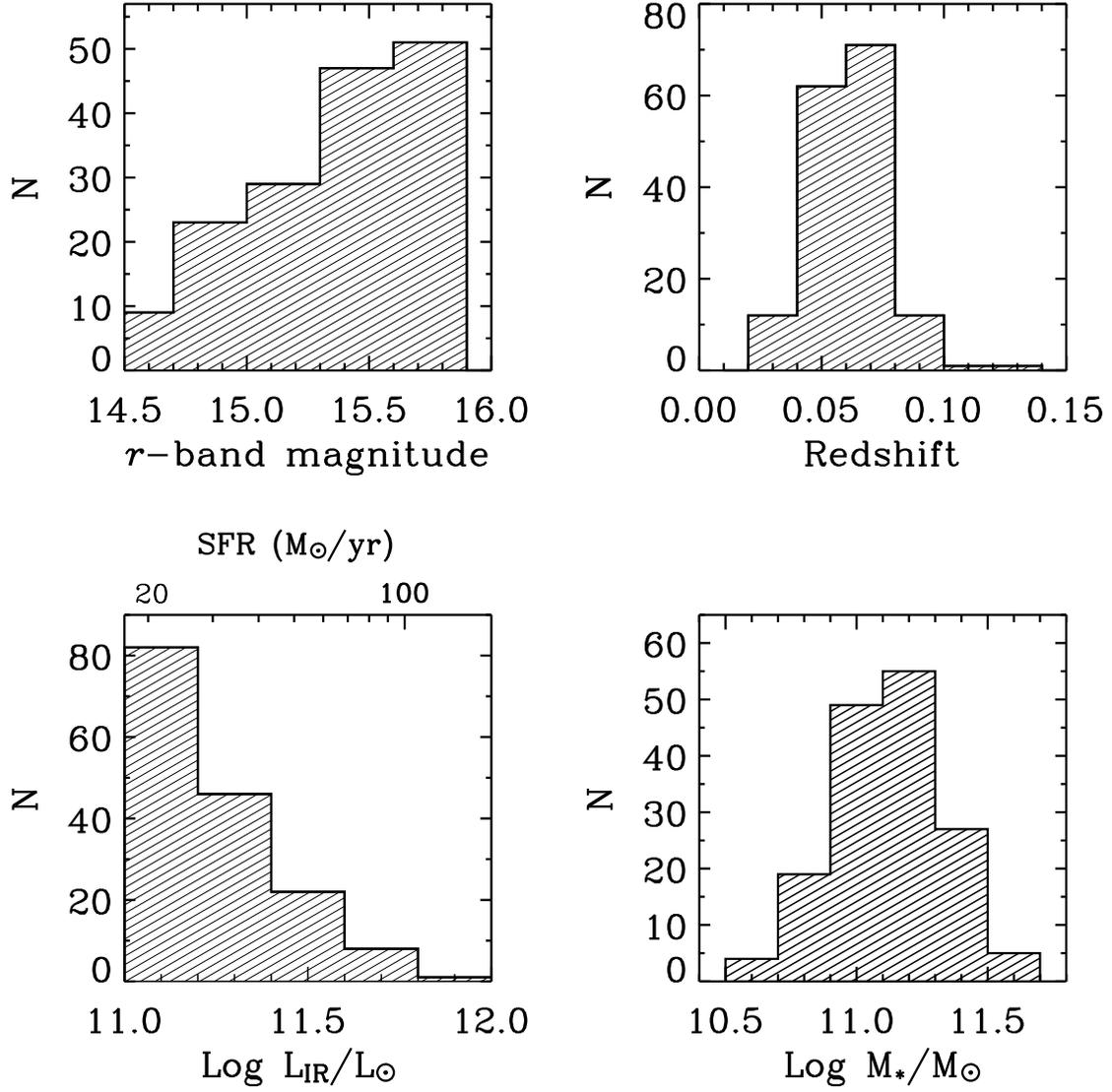}
\caption{$r$-band magnitude, redshift, infrared luminosity and stellar mass
distributions for our luminous infrared galaxies (LIRGs) sample. The
corresponding SFR scale is shown in the top horizontal axis in the
bottom left panel.
}
\end{figure}

\clearpage

\begin{figure}
\figurenum{2}
\epsscale{}
\plotone{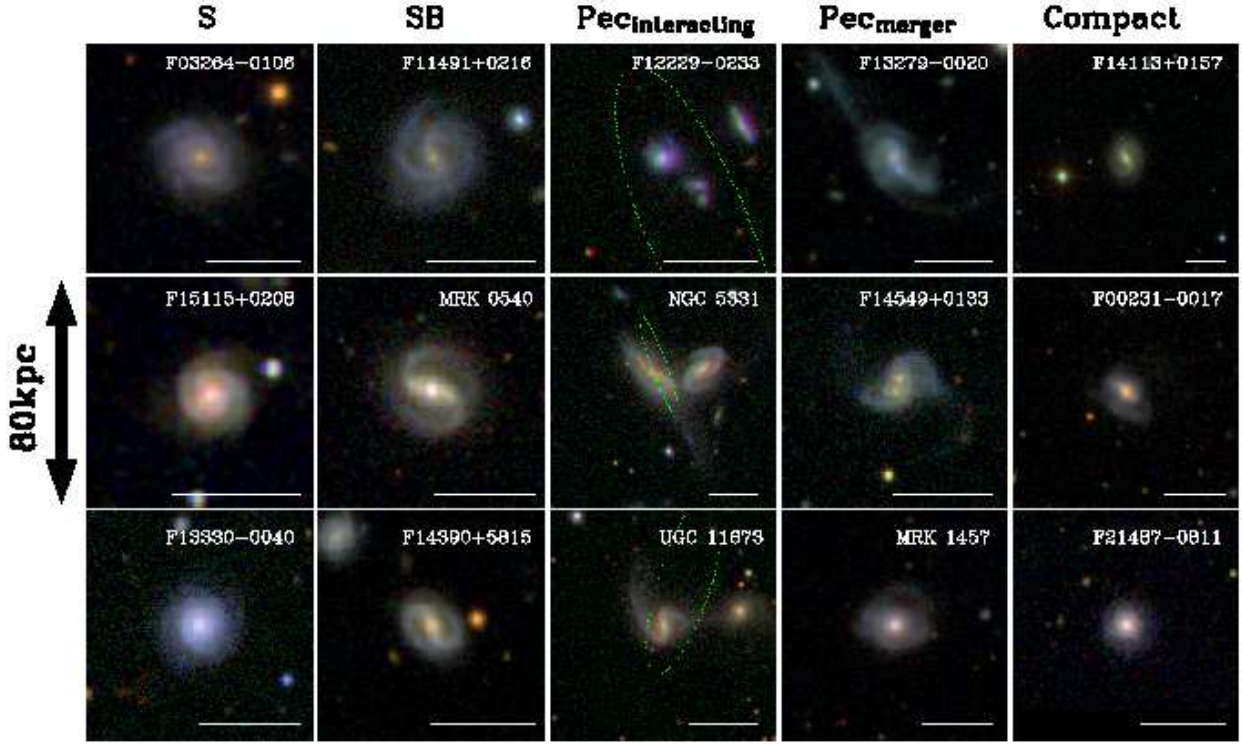}
\caption{
Three color images of 15 galaxies. Each color image is produced
by combining three ($g$, $r$, $i$) SDSS images using the method 
of Lupton et al. (2004). The images are grouped 
	into five morphological types: S - nonbarred spiral, SB - barred galaxies, 
	Pec$_{\rm interacting}$ - interacting galaxies, Pec$_{\rm merger}$-
merging galaxies, and Compact galaxies. Three examples are shown for
each morphology type. The size of each image is chosen to be 80 kpc at the
redshift of our sample galaxy. A horizontal bar in each panel indicates
an angular scale of 30$\arcsec$. The IRAS $2\sigma$ error ellipse is shown
for the three galaxies in the middle column.
}
\end{figure}

\clearpage

\begin{figure}
\figurenum{3}
\epsscale{}
\plotone{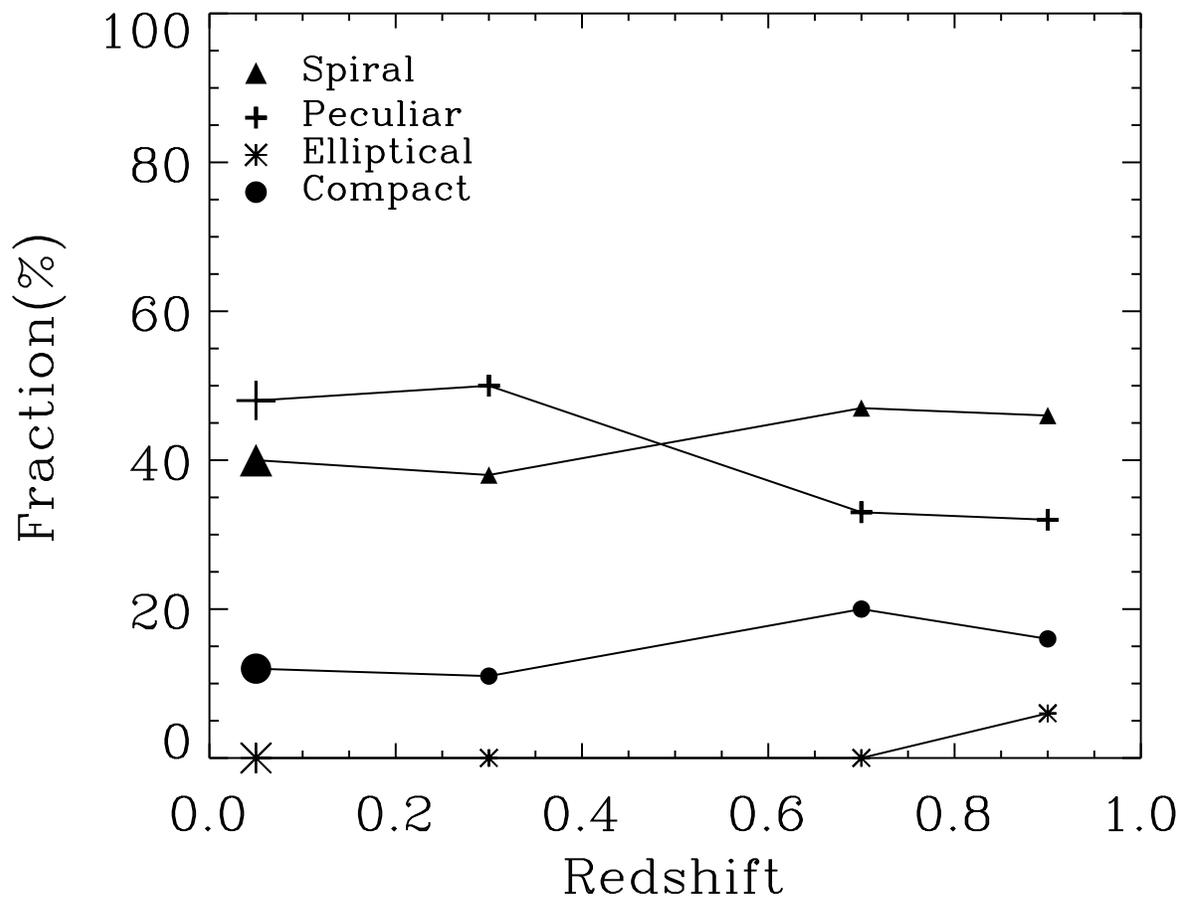}
\caption{ Morphological classification of LIRGs in the local universe
	(at $z\sim 0.05$, large symbols) compared with those of
Melbourne et al. (small symbol) at higher redshift.
        The spiral category includes barred, nonbarred and highly-inclined spirals. In
	the local sample, there are no ellipticals. The redshift shown
	is the median value of galaxy redshifts in each bin. 
}
\end{figure}

\begin{figure}
\figurenum{4}
\epsscale{}
\plotone{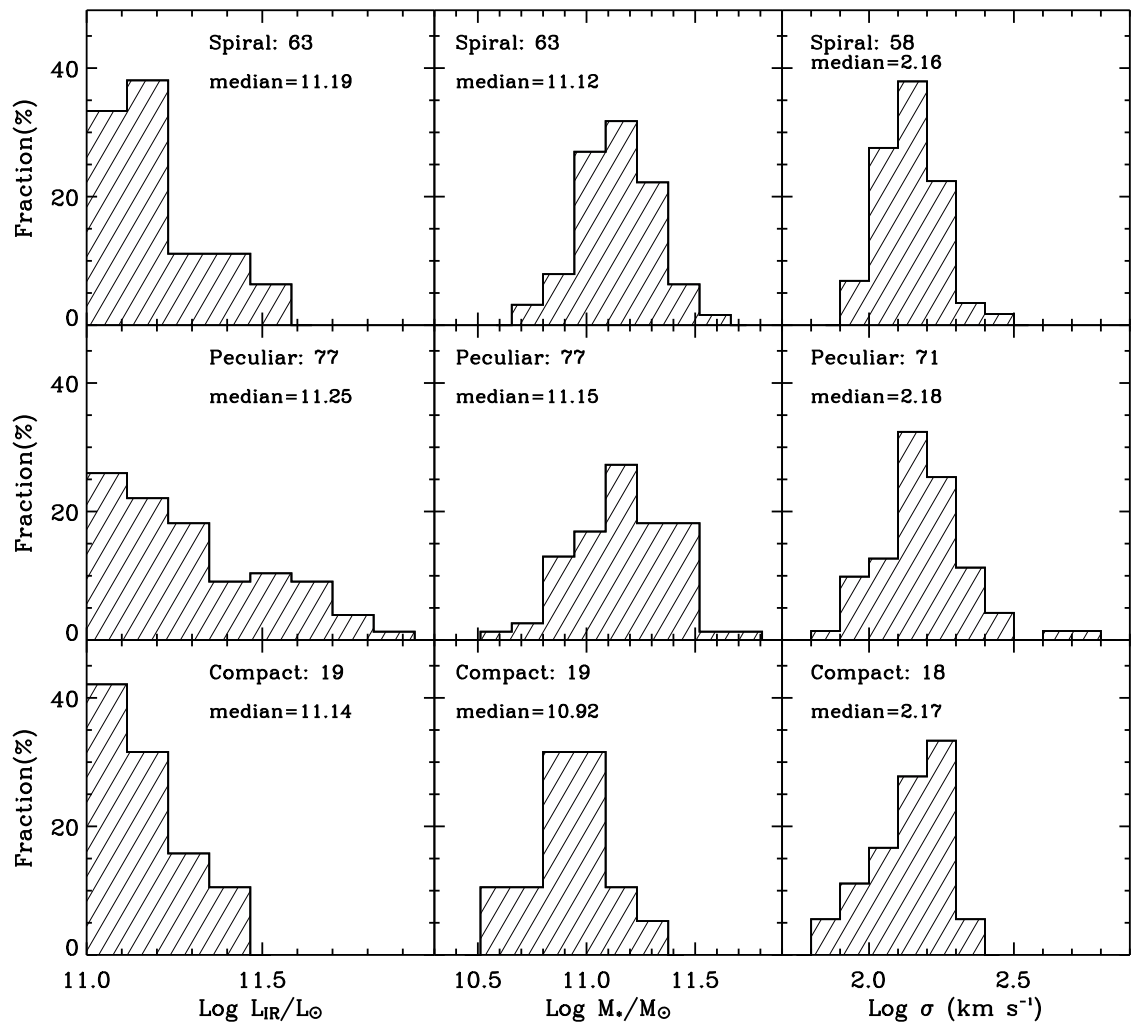}
\caption{Distribution of infrared luminosity, stellar mass and velocity dispersion for
different morphological types of local LIRGs.}
\end{figure}

\clearpage

\begin{figure}
\figurenum{5}
\epsscale{1.1}
\plottwo{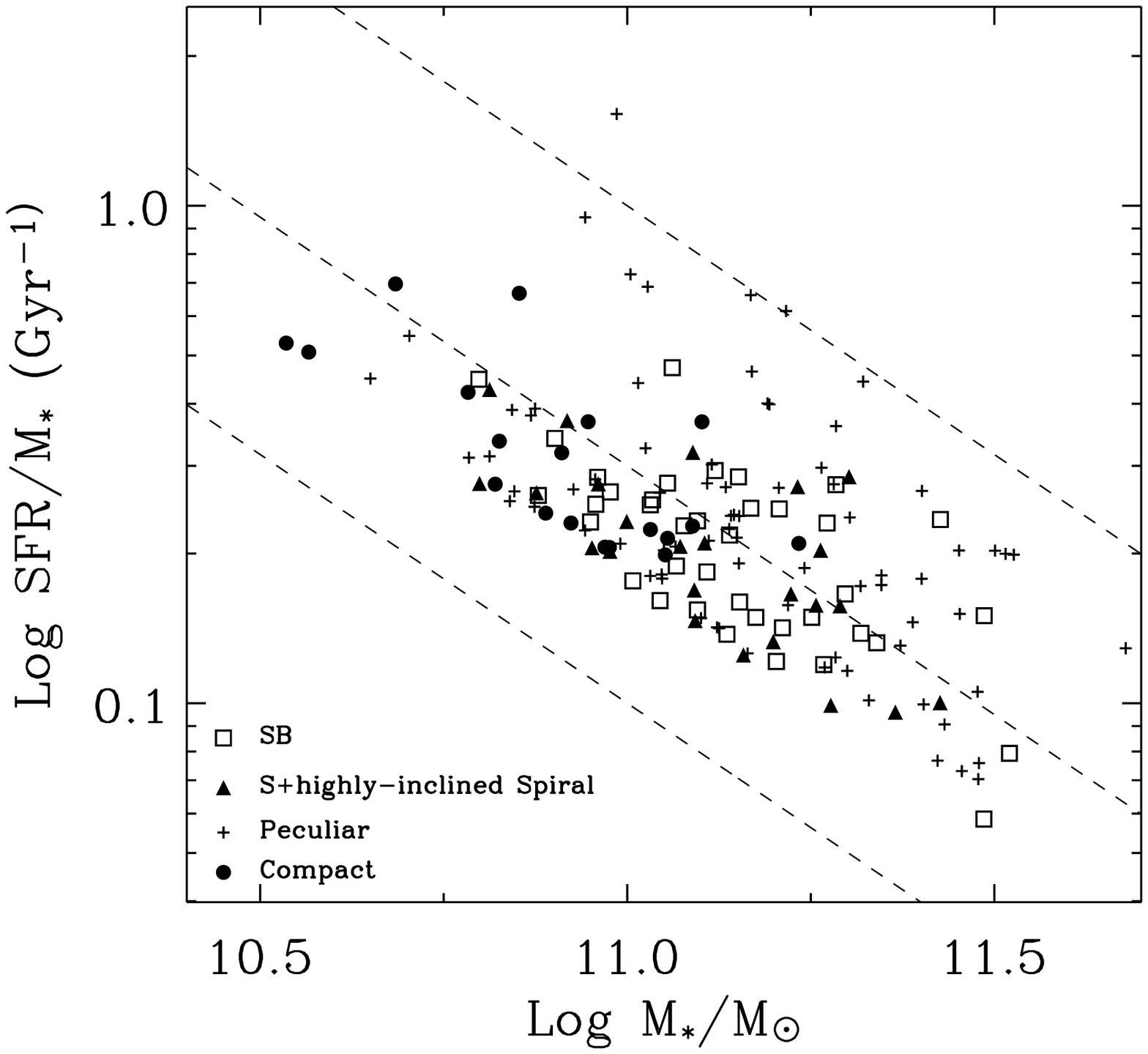}{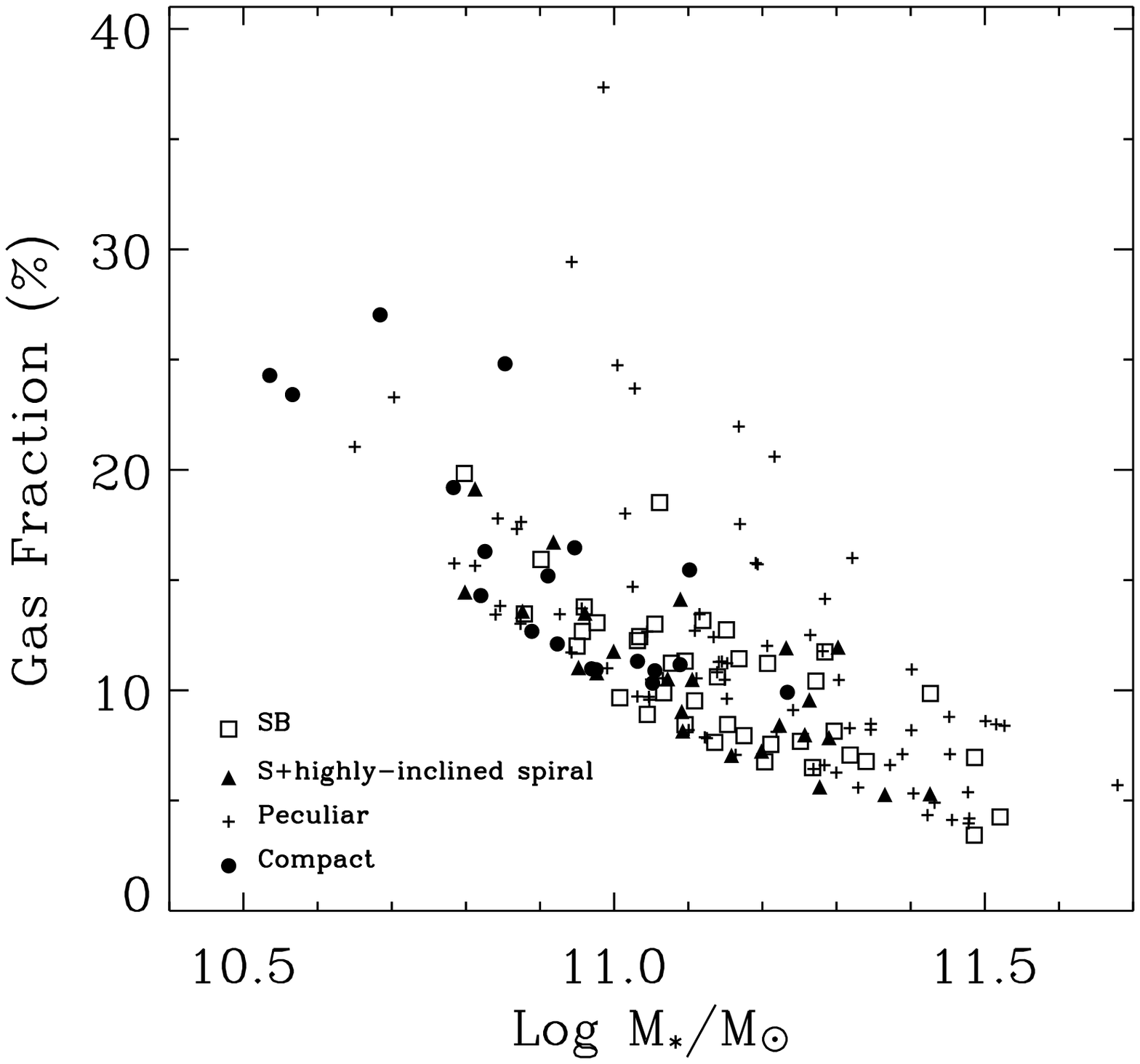}
\caption{The left panel shows the specific star formation rate (SFR per unit stellar mass) 
in units of Gyr$^{-1}$ vs. stellar mass for different morphological types of LIRGs. 
The three dashed lines (from bottom to top) denote a SFR  of 10, 30, and 100 $\myear$
respectively. The right panel shows the cold molecular mass fraction vs. stellar mass.
} 
\end{figure}

\end{document}